\documentclass[preprint,prb,aps,showpacs,showkeys,amsmath]{revtex4-2}
\usepackage{euscript}
\usepackage{graphics}
\usepackage{graphicx}
\usepackage{indentfirst}

\begin{document}

\noindent\fbox{
  \parbox{\textwidth}{
    This article may be downloaded for personal use only. Any other use
    requires prior permission of the author and AIP Publishing. This
    article appeared in ``Physics of Fluids 15, 3434–3442 (2003)",
    and may be found at https://doi.org/10.1063/1.1613648
  }
}

\title{Fourth-order statistical moments of the velocity gradient tensor in homogeneous, isotropic turbulence.}

\author{Juan Hierro}
\affiliation{LITEC, Consejo Superior de Investigaciones
Cient\'{\i}ficas\\
Mar\'{\i}a de Luna 3, Zaragoza 50018, Spain}
\author{C\'{e}sar Dopazo}
\affiliation{CIEMAT, Centro de Investigaciones Energ\'{e}ticas, MedioAmbientales y Tecnol\'{o}gicas \\
Avda. Complutense, 22, Madrid 28040, Spain}

\date{\today}

\begin{abstract}
A compact expression of fourth-order statistical moments of the velocity gradient tensor in homogeneous, isotropic, incompressible turbulence is obtained as a function of its invariants and of generic components of the velocity gradient. This single, compact expression is in full agreement with the four different expressions previously obtained by Siggia as functions of the same invariants and of generic components of the vorticity vector and the strain tensor; however, some discrepancies arise with respect to a similar, single expression obtained by Phan-Thien and Antonia. The used algorithm may be easily extended to handle higher order statistical moments of the velocity gradient.
\end{abstract}


\maketitle

\section{INTRODUCTION.}
\label{sec:1}

An expression for the second-order statistical moments of the velocity gradient tensor can be found in Hinze~\cite{Hinze:1975}. The naming convention $\mathbf{g}$, with $g_{ij} = u_{i,j}$ in covariant notation ($u_i$ is the $i$th component of the velocity vector field), will be used from now on to refer to the velocity gradient tensor. Champagne~\cite{Champagne:1978} calculated its third-order statistical moments. There are two degrees of freedom in these second-order and third-order statistical moments by isotropy, but only one is left in both situations after homogeneity is used. Their values may be related to the eigenvalues~\cite{Betchov:1956} of $\mathbf{g}$.

Later on, Siggia~\cite{Siggia:1981} proposed an algorithm to calculate the number of invariants which characterize the $n$th-order statistical moments of $\mathbf{g}$ in homogeneous, isotropic, incompressible turbulence. Moreover, this algorithm was particularized for $n=4$, to obtain the four invariants
\begin{align}
I_1 &= \langle (tr(\mathbf{s^2}))^2 \rangle = \langle (s_1^2+s_2^2+s_3^2)^2 \rangle \label{eq:C3} \\
I_2 &= \langle \omega^2 tr(\mathbf{s^2}) \rangle = \langle (\omega_1^2 + \omega_2^2 + \omega_3^2) (s_1^2 + s_2^2 + s_3^2) \rangle \notag \\
I_3 &= \langle \omega_i s_{ij} s_{jk} \omega_k \rangle =\langle \omega_1^2 s_1^2 + \omega_2^2 s_2^2 + \omega_3^2 s_3^2 \rangle \notag \\
I_4 &= \langle (\omega^2)^2 \rangle = \langle (\omega_1^2 + \omega_2^2 +  \omega_3^2)^2 \rangle \notag
\end{align}
where $\omega$ is the module of the vorticity vector, $\mathbf{s}$ is the strain tensor, $tr(\mathbf{s}^2)$ is the trace of the contracted product of $\mathbf{s}$ with itself (that is to say, $tr(\mathbf{s^2} = s_{ij} s_{ji}$), $\{s_1, s_2, s_3 \}$ are the eigenvalues of $\mathbf{s}$ and $\{\omega_1, \omega_2, \omega_3 \}$ are the components of the vorticity vector in the base of eigenvectors of $\mathbf{s}$. The brackets, $<>$, express a statistical average.

The objective of this paper is to compute a generic expression of the fourth-order statistical moments of $\mathbf{g}$ as a function of the fourth-order invariants, $I_{\alpha}, (\alpha=1,...,4)$, by means of an algorithm which could be easily extended to work with $n$th-order statistical moments.

It must be noticed that Siggia~\cite{Siggia:1981} and Phan-Thien and Antonia~\cite{Phan-Thien/Antonia:1994} also tackled this problem. 

From a practical point of view, the main difference with regard to the first reference~\cite{Siggia:1981} is that it expresses the fourth-order correlations of $\mathbf{g}$ in terms of its symmetric and antisymmetric parts by means of four different formulas \footnote{One formula with the fourth-order strain tensor only, another one with the fourth-order vorticity vector only and the two remaining formulas with fourth-order crossed terms between strain and vorticity.} instead of one compact expression which only depends on the the components of $\mathbf{g}$. Both approaches lead to the same results, the difference lies in the amount of calculations which are needed in the processing of the available information; Siggia's approach is better fit for working with vorticity and strain components whereas the approach of this paper is better fit for working with components of $\mathbf{g}$. Since the strain tensor and the vorticity vector are calculated from the symmetric and antisymmetric parts of $\mathbf{g}$, respectively, there are two advantages in the use of components of $\mathbf{g}$: first, one can obtain partial results about fourth-order correlations with fewer data (different components of $\mathbf{g}$) and, last, experimental errors are not added as it happens when the symmetric and antisymmetric parts are calculated as derived quantities. From a theoretical point of view, the deduction is widely different: no Gaussian ensemble is assumed at any point, the effect of three-dimensionality is taken into account explicitly, what proves the irrelevance of homogeneity in a different way, and, especially, the effect of the symmetries is explained through algorithms which are thought to be computationally performed.

On the other hand, the differences with respect to the second reference~\cite{Phan-Thien/Antonia:1994} are deeper, since it assumes that the fourth-order invariants, $I_{\alpha}, (\alpha=1,...,4)$, are mutually related.

\section{STATISTICAL MOMENTS OF THE VELOCITY GRADIENT TENSOR IN AN ISOTROPIC FIELD.}
\subsection{Effect of commutativity.}
\label{sec:2}

The value of the generic expression $\langle g_{ij} g_{kl} g_{mn} g_{pq} \rangle$ must be invariant under permutations of the components of $\mathbf{g}$ because of the commutativity of the product of real numbers and because all of these components have the same tensorial behaviour (second order) on their own. In other words, it must remain invariant under the $4! = 24$ permutations of pairs of indices which belong to the same gradient component. In general, any $n$th-order statistical moment of the velocity gradient must remain invariant under the $n!$ permutations of pairs of indices which belong to the same tensor component of $\mathbf{g}$.

\subsection{Effect of isotropy.}
\label{sec:3}

Although isotropy (invariance under rotation) is usually more restrictive than homogeneity (invariance under translation), it is convenient to consider its effect firstly. In an isotropic situation~\cite{Weyl:1939}, the $n$th-order statistical moments of $\mathbf{g}$ behave as components of an even order (2n) tensor which can be expressed as a linear combination of Kronecker delta products. This simplifies extraordinarily the analysis since each of these products corresponds to an arbitrary partition of the $2n$ indices of the $n$th-order statistical moment into pairs. Namely, the $n$th-order statistical moment of $\mathbf{g}$ can be expressed as a linear combination of $(2 n-1)!! = 1 \cdot 3 \cdot 5 \cdot ... \cdot (2 n - 1)$~\cite{Kearsley/Fong:1975,Phan-Thien/Antonia:1994} of such terms.

In Appendix~\ref{sec:A}, an algorithm which computes all the possible partitions into pairs of an even set of indices is explained and is extended to include the effect of commutativity which implies that the equivalent components must share the same numerical coefficient. The result for $n=4$ is

\begin{equation}
 \begin{split}
  \langle &u_{i,j}u_{k,l}u_{m,n}u_{p,q} \rangle = a_4 \delta_{ij} \delta_{kl} 
  \delta_{mn} \delta_{pq} +
  b_4 (\delta_{ij} \delta_{kl} \delta_{mp} \delta_{nq} + 
  \delta_{ij} \delta_{kp} \delta_{lq} \delta_{mn} + \delta_{ij} \delta_{km}
  \delta_{ln} \delta_{pq} +\\ 
 &\delta_{ip} \delta_{jq} \delta_{kl} \delta_{mn} +
  \delta_{im} \delta_{jn} \delta_{kl} \delta_{pq} + \delta_{ik} \delta_{jl}
  \delta_{mn} \delta_{pq}) + c_4 (\delta_{ij}\delta_{kl}\delta_{mq}\delta_{np}+
  \delta_{ij} \delta_{kq} \delta_{lp} \delta_{mn} + \delta_{ij} \delta_{kn}
  \delta_{lm} \delta_{pq} +\\ 
 &\delta_{iq} \delta_{jp} \delta_{kl} \delta_{mn} +
  \delta_{in} \delta_{jm} \delta_{kl} \delta_{pq} +
  \delta_{il} \delta_{jk} \delta_{mn} \delta_{pq}) +
  d_4 (\delta_{ij} \delta_{kn} \delta_{lp} \delta_{mq} +
  \delta_{ij} \delta_{kq} \delta_{lm} \delta_{np} + \delta_{in} \delta_{jp}
  \delta_{kl} \delta_{mq} +\\
 &\delta_{iq} \delta_{jm} \delta_{kl} \delta_{np} +
  \delta_{il} \delta_{jp} \delta_{kq} \delta_{mn} + \delta_{iq} \delta_{jk}
  \delta_{lp} \delta_{mn} + \delta_{il} \delta_{jm} \delta_{kn} \delta_{pq} +
  \delta_{in} \delta_{jk} \delta_{lm} \delta_{pq}) + e_4
 (\delta_{ij} \delta_{kn} \delta_{lq} \delta_{mp} +\\
 &\delta_{ij} \delta_{kq} \delta_{ln} \delta_{mp} +
  \delta_{ij} \delta_{km} \delta_{lq} \delta_{np} + \delta_{ij} \delta_{km}
  \delta_{lp} \delta_{nq} + \delta_{ij} \delta_{kp} \delta_{ln} \delta_{mq} +
  \delta_{ij} \delta_{kp} \delta_{lm} \delta_{nq} + \delta_{in} \delta_{jq}
  \delta_{kl} \delta_{mp} +\\
 &\delta_{iq} \delta_{jn} \delta_{kl} \delta_{mp} +
  \delta_{im} \delta_{jp} \delta_{kl} \delta_{nq} + \delta_{im} \delta_{jq}
  \delta_{kl} \delta_{np} + \delta_{ip} \delta_{jm} \delta_{kl} \delta_{nq} +
  \delta_{ip} \delta_{jn} \delta_{kl} \delta_{mq} +
  \delta_{il} \delta_{jq} \delta_{kp} \delta_{mn} +\\
 &\delta_{iq} \delta_{jl} \delta_{kp} \delta_{mn} +
  \delta_{ik} \delta_{jq} \delta_{lp} \delta_{mn} + \delta_{ik} \delta_{jp}
  \delta_{lq} \delta_{mn} + \delta_{ip} \delta_{jl} \delta_{kq} \delta_{mn} +
  \delta_{ip} \delta_{jk} \delta_{lq} \delta_{mn} + \delta_{il} \delta_{jn}
  \delta_{km} \delta_{pq} +\\
 &\delta_{in} \delta_{jl} \delta_{km} \delta_{pq} +
  \delta_{ik} \delta_{jn} \delta_{lm} \delta_{pq} + \delta_{ik} \delta_{jm}
  \delta_{ln} \delta_{pq} + \delta_{im} \delta_{jl} \delta_{kn} \delta_{pq} +
  \delta_{im} \delta_{jk} \delta_{ln} \delta_{pq}) + 
  f_4 (\delta_{ik}\delta_{jl}  \delta_{mp} \delta_{nq} +\\
 & \delta_{im} \delta_{jn} \delta_{kp} \delta_{lq} +
  \delta_{ip} \delta_{jq} \delta_{km} \delta_{ln}) + g_4 (
  \delta_{ik} \delta_{jl} \delta_{mq} \delta_{np} +
  \delta_{im} \delta_{jn} \delta_{kq} \delta_{lp} +
  \delta_{ip} \delta_{jq} \delta_{kn} \delta_{lm} + \delta_{iq} \delta_{jp}
  \delta_{km} \delta_{ln} +\\
 &\delta_{in} \delta_{jm} \delta_{kp} \delta_{lq} +
  \delta_{il} \delta_{jk} \delta_{mp} \delta_{nq}) + h_4
 (\delta_{il} \delta_{jk} \delta_{mq} \delta_{np} +
  \delta_{in} \delta_{jm} \delta_{kq} \delta_{lp} +
  \delta_{iq} \delta_{jp} \delta_{kn} \delta_{lm}) + \\
 &i_4 (\delta_{ik} \delta_{jq} \delta_{lm} \delta_{np} +
  \delta_{ik} \delta_{jn} \delta_{lp} \delta_{mq} +
  \delta_{ik} \delta_{jm} \delta_{lq} \delta_{np} + \delta_{ik} \delta_{jp}
  \delta_{ln} \delta_{mq} + 
  \delta_{im} \delta_{jl} \delta_{kq} \delta_{np} +
  \delta_{im} \delta_{jq} \delta_{kn} \delta_{lp} + \\
 &\delta_{im} \delta_{jp} \delta_{kq} \delta_{ln} +
  \delta_{im} \delta_{jk} \delta_{lp} \delta_{nq} +
  \delta_{ip} \delta_{jl} \delta_{kn} \delta_{mq} + \delta_{ip} \delta_{jn}
  \delta_{kq} \delta_{lm} + \delta_{ip} \delta_{jm} \delta_{kn} \delta_{lq} +
  \delta_{ip} \delta_{jk} \delta_{lm} \delta_{nq} +\\
 &\delta_{il} \delta_{jm} \delta_{kp} \delta_{nq} +
  \delta_{il} \delta_{jp} \delta_{km} \delta_{nq} +
  \delta_{il} \delta_{jn} \delta_{kq} \delta_{mp} + \delta_{il} \delta_{jq}
  \delta_{kn} \delta_{mp} + \delta_{in} \delta_{jp} \delta_{km} \delta_{lq} +
  \delta_{in} \delta_{jk} \delta_{lq} \delta_{mp} +\\
 &\delta_{in} \delta_{jl} \delta_{kp} \delta_{mq} +
  \delta_{in} \delta_{jq} \delta_{kp} \delta_{lm} +
  \delta_{iq} \delta_{jm} \delta_{kp} \delta_{ln} + \delta_{iq} \delta_{jk}
  \delta_{ln} \delta_{mp} + \delta_{iq} \delta_{jl} \delta_{km} \delta_{np} +
  \delta_{iq} \delta_{jn} \delta_{km} \delta_{lp}) + \\
 &j_4 (\delta_{ik} \delta_{jn} \delta_{lq} \delta_{mp} +
  \delta_{ik} \delta_{jq} \delta_{ln} \delta_{mp} +
  \delta_{im} \delta_{jq} \delta_{kp} \delta_{ln} + \delta_{im} \delta_{jl}
  \delta_{kp} \delta_{nq} + \delta_{ip} \delta_{jn} \delta_{km} \delta_{lq} +
  \delta_{ip} \delta_{jl} \delta_{km} \delta_{nq}) + \\
 &k_4 (\delta_{ik} \delta_{jm} \delta_{lp} \delta_{nq} +
  \delta_{ik} \delta_{jp} \delta_{lm} \delta_{nq} + 
  \delta_{im} \delta_{jp} \delta_{kn} \delta_{lq} + \delta_{im} \delta_{jk} 
  \delta_{lq} \delta_{np} + \delta_{ip} \delta_{jm} \delta_{kq} \delta_{ln} +
  \delta_{ip} \delta_{jk} \delta_{ln} \delta_{mq} + \\
 &\delta_{il} \delta_{jq} \delta_{km} \delta_{np} +
  \delta_{il} \delta_{jn} \delta_{kp} \delta_{mq} +
  \delta_{in} \delta_{jl} \delta_{kq} \delta_{mp} +
  \delta_{in} \delta_{jq} \delta_{km} \delta_{lp} +
  \delta_{iq} \delta_{jl} \delta_{kn} \delta_{mp} +
  \delta_{iq} \delta_{jn} \delta_{kp} \delta_{lm}) + \\
 &l_4 (\delta_{il} \delta_{jm} \delta_{kq} \delta_{np} +
  \delta_{il} \delta_{jp} \delta_{kn} \delta_{mq} +
  \delta_{in} \delta_{jk} \delta_{lp} \delta_{mq} + \delta_{in} \delta_{jp}
  \delta_{kq} \delta_{lm} +
  \delta_{iq} \delta_{jk} \delta_{lm} \delta_{np} +
  \delta_{iq} \delta_{jm} \delta_{kn} \delta_{lp})
 \end{split} \label{eq:C6}
\end{equation}

\section{ADDITIONAL RESTRICTIONS ON EQ.(\ref{eq:C6}).}
\subsection{Restrictions due to incompressibility.}
\label{sec:4}

Incompressibility implies that the contraction of a pair of indices pertaining to the same velocity gradient (contractions of $i$ with $j$, of $k$ with $l$, of $m$ with $n$ or of $p$ with $q$) in Eq.(\ref{eq:C6}) yields no contribution. For instance, 
\begin{equation}
\langle g_{ij} g_{kl} g_{mn} g_{pq} \rangle \delta_{pq} = 0
\label{eq:C7}
\end{equation}
Eq.(\ref{eq:C7}) is expanded by means of an algorithm explained in Appendix~\ref{sec:B}. As a result, the following five restrictions must be satisfied by the numerical coefficients of Eq.(\ref{eq:C6})
\begin{align}
&3 a_4 + 3 b_4 + 3 c_4 = a_4 + b_4 + c_4 = 0 \label{eq:C8} \\
&3 b_4 + 4 e_4 + f_4 + g_4 = 0 \notag \\
&3 c_4 + 2 d_4 + 2 e_4 + g_4 + h_4 = 0 \notag \\
&3 d_4 + 3 i_4 + 3 l_4 = d_4 + i_4 + l_4 = 0 \notag \\
&3 e_4 + 3 i_4 + j_4 + 2 k_4 = 0 \notag
\end{align}

It should be noticed that the contractions of $i$ with $j$, of $k$ with $l$ and of $m$ with $n$ do not modify the incompressible restrictions given by Eqs.(\ref{eq:C8}), although Eq.(\ref{eq:C7}) is certainly modified ($\delta_{pq}$ is replaced by either $\delta_{ij}$ or $\delta_{kl}$ or $\delta_{mn}$).

\subsection{Restriction due to homogeneity.}
\label{sec:5}

To impose the homogeneity condition on Eq.(\ref{eq:C6}), one may start by rewriting
\begin{equation}
\langle g_{ij} g_{kl} g_{mn} g_{pq} \rangle = \langle u_{i,j} u_{k,l} u_{m,n} u_p \rangle_{,q} - \langle u_{i,jq} u_{k,l} u_{m,n} u_p \rangle - \langle u_{k,lq} u_{i,j} u_{m,n} u_p \rangle - \langle u_{m,nq} u_{k,l} u_{i,j} u_p \rangle
\label{eq:C8a}
\end{equation}

Homogeneity implies, precisely, that the first term on the right-hand side of Eq.(\ref{eq:C8a}) vanishes. The remaining contributions are, except for the sign, a generic element of the tensor $\langle u_{i,jq} u_{k,l} u_{m,n} u_p \rangle$, plus a generic element of the same tensor where indices $i$ and $j$ have been exchanged with indices $k$ and $l$, respectively, plus a generic element of the same tensor where indices $i$ and $j$ have been exchanged with indices $m$ and $n$, respectively. Hence, $\langle u_{i,jq} u_{k,l} u_{m,n} u_p \rangle$ must be examined more carefully.

It is an eighth-order isotropic tensor which is the product of a generic component of the velocity Hessian tensor, times two generic components of $\mathbf{g}$, times a component of the velocity vector. Consequently, it is invariant under exchange of the indices which correspond to the two derivatives of the Hessian ($j \, \leftrightarrow \, q$, in the second term in the right-hand side of Eq.(\ref{eq:C8a})) and of the two pairs of indices which correspond to the two components of the velocity gradient ($k \,\leftrightarrow \, m$ and $l \, \leftrightarrow \, n$, in that term).

In Appendix~\ref{sec:C}, the commutativity invariance, which led to Eq.(\ref{eq:C6}) from a generic expression of an eighth-order isotropic tensor, is adapted to $\langle u_{i,jq} u_{k,l} u_{m,n} u_p \rangle$. The result is given by Eq.(\ref{eq:C9}) which is substituted in Eq.(\ref{eq:C8a}) together with alike expressions corresponding to the last two terms in the right-hand side of Eq.(\ref{eq:C8a}). The final result is that Eq.(\ref{eq:C8a}) is fulfilled if and only if
\begin{equation}
a_4 - 6 c_4 + 8 d_4 + 3 h_4 - 6 l_4 = 0
\label{eq:C20}
\end{equation}

Therefore, Eq.(\ref{eq:C20}) is an additional restriction due to the homogeneity condition which should be satisfied by the numerical coefficients of Eq.(\ref{eq:C6}).

\subsection{Recapitulation.}
\label{sec:6}

Because of isotropy and commutativity, the fourth-order moment of $\mathbf{g}$ in Eq.(\ref{eq:C6}) possesses twelve degrees of freedom. After applying the incompressibility and homogeneity conditions, Eqs.(\ref{eq:C8}) and (\ref{eq:C20}), the degrees of freedom are reduced to six.

However, it has been stated in Section~\ref{sec:1} that the fourth-order statistical moment of $\mathbf{g}$ should be characterized by only four invariants. Moreover, Siggia~\cite{Siggia:1981} proved that homogeneity has no effect on the fourth-order statistical moment of $\mathbf{g}$.

To solve the previous contradiction, it must be additionally considered that one is dealing with a three-dimensional space. This restricts the number of numerical coefficients which are measurable, as it will be explained in the following Subsection.

\subsection{Effect of three-dimensionality on third-order statistical moments.}
\label{sec:7}

Fourth-order statistical moments of $\mathbf{g}$ contain four pairs of indices; however, in a three-dimensional world, there will be, at most, three different indices. In other words, one cannot measure the full tensor given by Eq.(\ref{eq:C6}), but only those combinations of coefficients which correspond to the presence of, at least, four repeated indices. It should be remembered that Eq.(\ref{eq:C6}) implies that all statistical moments with an odd number of repeated indices are zero because of the properties of the Kronecker delta.

In practice, the measurable quantities correspond to breakdowns of the total number of indices, $2n$, into the addition of either three even numbers, or two even numbers, or just the $2n$ number itself. If $n=4$, measurable quantities correspond to either generic situations where there are $\{4,2,2\}$ repeated indices, or $\{4,4\}$ repeated indices, or $\{6,2\}$ repeated indices, or $\{8\}$ repeated indices. However, combinations of coefficients corresponding to situations with only one or two repeated indices may be expressed as equivalent combinations of situations with three repeated indices (for instance, $iiiijjjj = iiiijjkk + iiiijkjk + iiiijkkj$). If $n=4$, only situations of the kind $\{4,2,2\}$ must be studied in order to find all the measurable combinations of coefficients of Eq.(\ref{eq:C6}).

From now on, only $n=4$ is considered.

The number of components of the fourth-order statistical moment of the velocity gradient with four repeated indices is that of the combinations of eight elements taken four by four, namely $70$. However, not all of them must be studied. In effect, one may apply commutativity to build classes of equivalence within the $70$ components. If one writes down the four indices which are repeated to symbolize each component of the kind $\{4,2,2\}$, one gets the following nine equivalence classes: \{ijkl, ijmn, ijpq, klmn, klpq, mnpq\}, \{ijkm, ijkp, ijmp, iklm, iklp, klmp, ikmn, ipmn, kmnp, ikpq, impq, kmpq\}, \{ijln, ijlq, ijnq, jkln, jklq, klnq, jlmn, jmnq, lmnq, jlpq, jnpq, lnpq\}, \{ijkn, ijkq, ijmq, ijlm, ijlp, ijnp, ikln, iklq, klmq, jklm, jklp, klnp, ilmn, imnq, kmnq, jkmn, jmnp, lmnp, ilpq, inpq, knpq, jkpq, jmpq, lmpq\}, \{ikmp\}, \{jlnq\}, \{ikmq, iknp, ilmp, jkmp\}, \{ilnq, jknq, jlmq, jlnp\}, \{iknq, ilmq, ilnp, jkmq, jknp, jlmp\}. Thus, only one member of each equivalence class must be studied.

By taking the first member of each class and writing down the corresponding three measurable moments of kind $\{4,2,2\}$ (for instance, with the first class: $(\delta_{ij} \delta_{kl} + \delta_{ik} \delta_{jl} + \delta_{il} \delta_{jk}) \delta_{mn} \delta_{pq}$, $(\delta_{ij} \delta_{kl} + \delta_{ik} \delta_{jl} + \delta_{il} \delta_{jk}) \delta_{mp} \delta_{nq}$, $(\delta_{ij} \delta_{kl} + \delta_{ik} \delta_{jl} + \delta_{il} \delta_{jk}) \delta_{mq} \delta_{np}$), one finally gets that all measurable fourth-order statistical moments of $\mathbf{g}$ are expressible as linear combinations of the following twelve combinations of coefficients
\begin{alignat}{3}
a_4&+b_4+c_4, & \qquad b_4&+2 e_4, & \qquad b_4&+f_4+g_4, \label{eq:C30} \\
c_4&+d_4+e_4, & \qquad c_4&+g_4+h_4, & \qquad d_4&+i_4+l_4, \notag \\
e_4&+i_4+j_4, & \qquad e_4&+i_4+k_4, & \qquad f_4&+2 j_4, \notag \\
f_4&+ 2 k_4, & \qquad g_4&+ 2 i_4, & \qquad h_4&+k_4+l_4. \notag
\end{alignat}
However, it is immediate to realize that there are only nine linearly independent combinations in Eqs.(\ref{eq:C30}), as
\begin{align}
&(f_4 + 2 k_4) + (g_4 + 2 i_4) + (b_4 + 2 e_4) = 2 (e_4 + i_4 + k_4) +
(b_4 + f_4 + g_4) \label{eq:C31} \\
&(f_4 + 2 j_4) + (g_4 + 2 i_4) + (b_4 + 2 e_4) = 2 (e_4 + i_4 + j_4) +
(b_4 + f_4 + g_4) \notag \\
&(c_4 + d_4 + e_4) + (g_4 + 2 i_4) + (h_4 + k_4 + l_4) = (c_4 + g_4 + h_4) +
(d_4 + i_4 + l_4) + (e_4 + i_4 + k_4) \notag
\end{align}

Moreover, incompressibility restrictions, Eqs.(\ref{eq:C8}), translate straightforwardly into the combinations of coefficients given by Eqs(\ref{eq:C30}).
\begin{align}
&a_4 + b_4 + c_4 = 0 \label{eq:C32} \\
&d_4 + i_4 + l_4 = 0 \notag \\
&3 b_4 + 4 e_4 + f_4 + g_4 = 2 (b_4 + 2 e_4) + (b_4 + f_4 + g_4) = 0 \notag \\
&3 c_4 + 2 d_4 + 2 e_4 + g_4 + h_4 = 2 (c_4+d_4+e_4) + (c_4+g_4+h_4) = 0 
\notag \\
&3 e_4 + 3 i_4 + j_4 + 2 k_4 = 2 (e_4 + i_4 + k_4) + (e_4 + i_4 + j_4) = 0
\notag
\end{align}

In summary, Eqs.(\ref{eq:C31}) and (\ref{eq:C32}) only leave four degrees of freedom out of the twelve combinations of coefficients which appeared to be relevant at first. These four degrees of freedom correspond to the four invariants given by Eqs.(\ref{eq:C3}). One could be tempted to think that the homogeneity restriction removes an additional degree of freedom, establishing a relation among the four invariants; however, there is no linear combination of the combinations of coefficients given by Eqs.(\ref{eq:C31}), reproducing the homogeneity restriction, Eq.(\ref{eq:C20}). That is to say, in three-dimensions, the homogeneity restriction on the fourth-order statistical moment of $\mathbf{g}$ becomes irrelevant since it cannot be measured.

\section{NUMERICAL COEFFICIENTS OF THE GENERIC EXPRESSION OF THE FOURTH-ORDER STATISTICAL MOMENT.}
\label{sec:8}

In this Section, some possible values for the numerical coefficients in Eq.(\ref{eq:C6}), are obtained.

In the first place, the following components of the fourth-order statistical moment of the velocity gradient
\begin{align}
F_1 &= \langle u_{1,1}^4 \rangle = 4 I_1/105 \label{eq:C24} \\
F_2 &= \langle u_{1,1}^2 u_{2,1}^2 \rangle = I_1 / 105 + I_2 / 70 - I_3 / 105 
\notag \\
F_3 &= \langle u_{2,1}^4 \rangle = 3 I_1/140 + 11 I_2/140 - 3 I_3/35 + I_4/80 
\notag \\
F_4 &= \langle u_{1,1}^2 u_{2,3}^2 \rangle = I_1/105+I_2/210+2 I_3/105 \notag
\end{align}
will be considered, instead of the four invariants given by Eqs.(\ref{eq:C3}). These components were introduced in Ref. 4 because $F_1, F_2$ and $F_3$ were specially suitable to be experimentally measured by means of crossed wires. However, most of the reviewed experimental measurements~\cite{VanAtta/Antonia:1980,Phan-Thien/Antonia:1994,Water:1996} only concern longitudinal and transverse components of the velocity gradient; that is to say, $F_1$ and $F_3$ are the only components which may be obtained from them. It must be mentioned that there are experimental measurements of the fourth-order invariants, $\{I_1, I_2, I_3, I_4\}$, at a moderate Reynolds number in Ref. 11. In Appendix~\ref{sec:E}, Eqs.(\ref{eq:C24}) will be derived as an example of the rather lengthy, though straightforward, calculations which are involved in going from a formulation which uses components of $\mathbf{g}$ into a formulation which uses components of strain and vorticity.

In the second step, it is proved in Appendix~\ref{sec:D} that all measurable statistical moments may be expressed as combinations of $(b_4+2e_4)$, $(c_4 + d_4 + e_4)$, $(g_4 + 2 i_4)$ and $(e_4 + i_4 + j_4)$ which, in their turn, may be written as functions of the components of $\mathbf{g}$ given by Eq.(\ref{eq:C24}).
\begin{align}
b_4 &+ 2 e_4 = -F_4 / 2 \label{eq:C27} \\
c_4 &+ d_4 + e_4 = F_4 / 2 - F_1 / 4 \notag \\
e_4 &+ i_4 + j_4 = F_2 - F_4 \notag \\
g_4 &+ 2 i_4 = 2 F_2 - F_4 / 2 - F_3 / 3 \notag
\end{align}

In the final step, Eqs.(\ref{eq:C30}) must be inverted in order to obtain the coefficients of Eq.(\ref{eq:C6}) as functions of measurable quantities. In this inversion, there are three degrees of freedom left; one of them may be canceled by applying the homogeneity condition, Eq.(\ref{eq:C20}). The other two may be canceled by making $i_4 = l_4 = 0$; this, because of incompressibility (Eqs.(\ref{eq:C8})), also implies that $d_4 = 0$. After the inversion and taking into account Eqs.(\ref{eq:C25}) and (\ref{eq:C27}), one finally gets
\begin{align}
a_4 &= -7 F_1 / 8 + 9 F_2 / 4 - 3 F_3 / 8 + 3 F_4 / 2 \label{eq:C28} \\
b_4 &=  3 F_1 / 4 - 3 F_2 / 2  +  F_3 / 4 - 3 F_4 / 2 \notag \\
c_4 &=    F_1 / 8 - 3 F_2 / 4  +  F_3 / 8 \notag \\
d_4 &= 0 \notag \\
e_4 &= -3 F_1 / 8 + 3 F_2 / 4  -  F_3 / 8  +  F_4 / 2 \notag \\
f_4 &= -3 F_1 / 4  -  F_2 / 2  +  F_3 / 12 + 3 F_4 \notag \\
g_4 &=                2 F_2   -   F_3 / 3  -  F_4 / 2 \notag \\
h_4 &=  3 F_1 / 8 - 5 F_2 / 4 + 5 F_3 / 24 -  F_4 / 2 \notag \\
i_4 &= 0 \notag \\
j_4 &=  3 F_1 / 8  +  F_2 / 4  +  F_3 / 8 - 3 F_4 / 2 \notag \\
k_4 &=  3 F_1 / 8 - 5 F_2 / 4  +  F_3 / 8 \notag \\
l_4 &= 0 \notag
\end{align}
or, alternatively, using the definitions in Eqs.(\ref{eq:C24})
\begin{align}
a_4 &= -19 I_1/3360 + 11 I_2/1120 + 11 I_3/280 - 3 I_4/640 \label{eq:C29} \\
b_4 &= 3 I_1 / 560 - I_2 / 112 - I_3 / 28 + I_4 / 320 \notag \\
c_4 &= I_1 / 3360 - I_2 / 1120 - I_3 / 280 + I_4 /640 \notag \\
d_4 &= 0 \notag \\
e_4 &= -17 I_1 / 3360 + 11 I_2 / 3360 + 11 I_3 / 840 - I_4 /640 \notag \\
f_4 &= - I_1 / 336 + 23 I_2 / 1680 + 23 I_3 / 420 + I_4 / 960 \notag \\
g_4 &= I_1 / 140 - I_4 / 240 \notag \\
h_4 &= I_1 / 480 - 13 I_2 / 3360 - 13 I_3 / 840 + I_4 / 384 \notag \\
i_4 &= 0 \notag \\
j_4 &= 17 I_1 / 3360 + I_2 / 160 - I_3 / 24 + I_4 / 640 \notag \\
k_4 &= 17 I_1 / 3360 - 9 I_2 / 1120 + I_3 / 840 + I_4 / 640 \notag \\
l_4 &= 0 \notag
\end{align}

It should be noticed that Eqs.(\ref{eq:C28}) and (\ref{eq:C29}) are not the only possibilities since, after applying the homogeneity condition, there were two degrees of freedom which were canceled by taking, arbitrarily, $i_4 = l_4 = 0$.

\section{CONCLUSIONS.}
\label{sec:9}

The main contribution of this paper is Eq.(\ref{eq:C6}). Together with Eqs.(\ref{eq:C28}) or (\ref{eq:C29}), it provides a full, compact form of the fourth-order statistical moments of $\mathbf{g}$ in a homogeneous, isotropic, incompressible flow. Any velocity gradient model must satisfy this formal expression which also serves as a test of the degree of homogeneity, isotropy and incompressibility of a given flow.

As examples of fourth order correlations which should be related to each other, in homogeneous, isotropic, incompressible situations, one may start with those given by Eqs.(\ref{eq:C24}). They should be independent of any particular direction; that is to say, the particular values of the repeated indices must not cause any difference, the correlations must depend only on their disposition. For example, $\langle u_{1,1}^4 \rangle = \langle u_{2,2}^4 \rangle = \langle u_{3,3}^4 \rangle$. However, the relations among the different measurable quantities, Subsection \ref{sec:7}, are more interesting because they are not trivial. Some examples are: $\langle u_{1,1}^2 u_{2,2} u_{3,3} \rangle = a_4 + b_4 + c_4 = 0$; $\langle u_{1,2}^2 u_{1,3}^2 \rangle = f_4 + 2 j_4 = F_3 / 3$; $\langle u_{1,2}^2 u_{3,1}^2 \rangle = f_4 + 2 k_4 = 3 (F_4 - F_2) + F_3 / 3$; $\langle u_{1,1}^2 u_{1,2}^2 \rangle = F_2 = \langle u_{1,1}^2 u_{2,1}^2 \rangle$; $\langle u_{1,1}^2 u_{2,3} u_{3,2} \rangle = c_4 + g_4 + h_4 = F_1 / 2 - F_4$.

The results obtained are in full agreement with those of Ref. 4. In particular, Eqs. (6), (7a), (7b) and (8) of Siggia~\cite{Siggia:1981} are equivalent to those derived from Eq.(\ref{eq:C6}) and Eqs.(\ref{eq:C29}) when $\mathbf{g}$ is broken down into the strain tensor and the vorticity vector.

However, there is no agreement with Ref. 5, although the algorithm explained there has been used here to build isotropic tensors of even order. The discrepancy arises from the use of Eq.(17) of Phan-Thien and Antonia~\cite{Phan-Thien/Antonia:1994} as a guide to build $\langle g_{ij} g_{kl} g_{mn} g_{pq} \rangle$ \footnote{The naming convention in~\cite{Phan-Thien/Antonia:1994} is slightly different; $\langle g_{im} g_{jn} g_{kp} g_{lq} \rangle$ is calculated instead of $\langle g_{ij} g_{kl} g_{mn} g_{pq} \rangle$.}. This Equation was previously obtained~\cite{Champagne:1978} for the second-order statistical moments of $\mathbf{g}$ and its use as a pattern to build fourth-order statistical moments of $\mathbf{g}$ is not justified. Were it valid, the following relation $F_1 = 3 F_2 / 2 = F_3 / 4 = 3 F_4 / 2$ (equivalent to $I_2 = I_3 / 3 = I_4 / 10 = 10 I_1 / 21$) should be satisfied~\cite{Phan-Thien/Antonia:1994}. Remembering the definitions $F_1 = \langle u_{1,1}^4 \rangle$ and $F_3 = \langle u_{1,2}^4 \rangle$ and that $\langle u_{1,2}^2 \rangle = 2 \langle u_{1,1}^2 \rangle$ in a homogeneous flow~\cite{Hinze:1975,Champagne:1978}, the flatness factor (fourth-order statistical moment, normalized by the square of the variance) of diagonal and non-diagonal components of $\mathbf{g}$ should be the same; however, neither DNS data~\cite{Vincent/Meneguzzi:1991}, nor experimental data~\cite{Tsinober/Kit/Dracos:1992} support this fact. Concretely, Eq.(17) of Ref. 5 is
\begin{equation*}
\langle g_{ij} g_{kl} \rangle = \frac{\langle g_{11}^2 \rangle}{2} (5 \delta_{ik} \delta_{jl} - 15 I^{(4)}_{ijkl} )
\end{equation*}
where $15 I^{(4)}_{ijkl} = \delta_{ij} \delta_{kl} + \delta_{ik} \delta_{jl} + \delta_{il} \delta_{jk}$. Problems arise when it is used as a pattern to build fourth-order correlations according to
\begin{equation*}
\begin{split}
\langle &g_{ij} g_{kl} g_{mn} g_{pq} \rangle = a (\delta_{ik} \delta_{jl} \delta_{mp} \delta_{nq} + \delta_{im} \delta_{jn} \delta_{kp} \delta_{lq} + \delta_{ip} \delta_{jq} \delta_{km} \delta_{ln}) + \\
& 15 b (\delta_{ik} \delta_{jl} I^{(4)}_{mnpq} + \delta_{im} \delta_{jn} I^{(4)}_{klpq} + \delta_{ip} \delta_{jq} I^{(4)}_{klmn} + \delta_{km} \delta_{ln} I^{(4)}_{ijpq} + \delta_{kp} \delta_{lq} I^{(4)}_{ijmn} + \delta_{mp} \delta_{nq} I^{(4)}_{ijkl}) + \\
&225 c (I^{(4)}_{ikjl} I^{(4)}_{mpnq} + I^{(4)}_{imjn} I^{(4)}_{kplq} + I^{(4)}_{ipjq} I^{(4)}_{kmln})
\end{split}
\end{equation*}
This Equation could serve to calculate all the components of the fourth-order correlation of $\mathbf{g}$; that is to say, each combination of Kronecker deltas in Eq.(\ref{eq:C6}). Nevertheless, it does not reproduce properly the symmetries between these components. After the application of incompressibility, there is only one degree of freedom instead of four.

To sum it up, the expression obtained here is more compact than that of Ref. 4, since there is no need to look for four different expressions in order to compute a generic fourth-order statistical moment of $\mathbf{g}$, and more accurate than that of Ref. 5, since relations among the invariants of the fourth-order statistical moment of $\mathbf{g}$ are not assumed.

\section*{ACKNOWLEDGEMENTS.}
Juan Hierro would like to thank the Spanish Ministry of Science and Technology for its support of this work through a FPU Fellowship.

\appendix

\section{ALGORITHM TO IMPOSE ISOTROPY AND COMMUTATIVITY.}
\label{sec:A}

To build an isotropic tensor of even order, the same algorithm as in Ref. 5 is used. It is based on the recursive nature of the partition of $2n$ indices into pairs. In effect, one may isolate one index as the first one and, then, build all the possible pairs of that first index with the rest of them; that is to say, one may build $(2 n - 1)$ pairs. The partitions into pairs of the preceding $2n$ indices are expressible as the product of each one of the preceding pairs times a partition into pairs of the remaining $(2 n - 2)$ indices. The recursion may be stopped when $n = 2$. At that point, there are three possible partitions into pairs of the four indices; for instance, $\{(ij,kl),(ik,jl),(il,jk)\}$, if each index is labeled: $i,j,k,l$.

The preceding algorithm is applied to $n=4$ in order to get the $105$ partitions into pairs (where each pair is equivalent to a Kronecker delta) of the $8$ indices which are labeled: $i,j,k,l,m,n,p,q$.

The next step is to look for the partitions into pairs which are equivalent due to the commutativity explained in Subsection~\ref{sec:2}. The equivalence classes are defined through the permutations of pairs of indices of a basic ordering which is taken as a reference. In the present situation, the basic ordering is the aforementioned $ijklmnpq$ and there are $4!=24$ permutations to consider.

The algorithm works in the following way. The first of the preceding $105$ partitions into pairs of the eight indices $ijklmnpq$ is subject to the action of the $24$ permutations and all the resulting partitions which are mutually different are included into the first class. Successive partitions are compared with all the elements which belong to some class. If there is a coincidence, that partition has been already taken into account; if not, one must build a new class as it was done with the first element. To compare different partitions, they must be first reordered. The used criterion to do that was, first, to reorder each pair so that its first element was less than the second one according to the basic ordering ($ijklmnpq$) from left to right and, last, to reorder the different pairs so that their first elements were in growing order according to the same basic ordering from left to right.

The final result is Eq.(\ref{eq:C6}) where the members of the same class of equivalence under commutativity share the same numerical coefficient.  

\section{ALGORITHM TO IMPOSE INCOMPRESSIBILITY.}
\label{sec:B}

To expand Eq.(\ref{eq:C7}), each element of Eq.(\ref{eq:C6}) is multiplied by $\delta_{pq}$. Then, the rules of contraction of Kronecker deltas are applied:

i) If two deltas have a common index, both deltas are replaced by one delta which is made up by the remaining two different indices; each one of them comes from a different original delta.

ii) If two deltas have both indices in common, both deltas are replaced by a numerical factor $3$.

In a final stage, the same reordering as in the Appendix~\ref{sec:A} is applied and all the coefficients which affect the same product of Kronecker deltas are added up. All these additions must be nil in order to keep incompressibility. The result is given by the Eqs.(\ref{eq:C8}).

\section{ALGORITHM TO IMPOSE HOMOGENEITY.}
\label{sec:C}

The application of the homogeneity condition to an $n$th-order statistical moment of the velocity gradient produces a linear combination of $n-1$ tensors. Each one of them is the statistical moment of one component of the velocity Hessian, times $n-2$ components of the velocity gradient, times one component of the velocity. The reason is immediate if one considers Eq.(\ref{eq:C8a}): a derivative-like component of one velocity gradient is isolated and applied to each one of the remaining $n-1$ velocity gradients in the original expression. These new tensors behave as the original ones as much as only isotropy is concerned. The difference comes from the set of indices permutations which leave them invariant.

A tensor so built remains invariant, because of commutativity, if the pairs of indices which correspond to velocity gradients are exchanged; that is to say, there are $(n-2)!$ permutations of indices which do not modify the tensor. Moreover, the derivative-like indices of the velocity Hessian are also exchangeable on their own due to the chain rule. Since there is only one Hessian component with two derivative-like indices, there are only $2! = 2$ allowed permutations of these kind of indices. Therefore, the total number of permutations that one must consider is $2 (n-2)!$. If $n=4$, the preceding quantity is equal to $4$ and all the permutations may be represented by $$\{ (),(j \, \leftrightarrow \, q),(k \, \leftrightarrow \, m, l \, \leftrightarrow \, n),(j \, \leftrightarrow \, q, k \, \leftrightarrow \, m, l \, \leftrightarrow \, n) \}$$ with regard to the second term in the right-hand side of Eq.(\ref{eq:C8a}).

By applying the same algorithm as in Appendix~\ref{sec:A}, with the previously stated four permutations of indices instead of the $24$ ones of the original version, the Equation (\ref{eq:C9}) is obtained as a generic expression of $\langle u_{i,jq} u_{k,l} u_{m,n} u_p \rangle$.

\begin{equation}
 \begin{split}
\langle &u_{i,jq}u_{k,l}u_{m,n}u_{p} \rangle = y_1 \delta_{ip} \delta_{jq} 
 \delta_{kl} \delta_{mn} +
 y_2 \delta_{ip} \delta_{jq} \delta_{km} \delta_{ln} + y_3
 \delta_{ip} \delta_{jq} \delta_{kn} \delta_{lm} + y_4 (\delta_{ij} \delta_{kl}
 \delta_{mn} \delta_{pq} + \\
&\delta_{iq} \delta_{jp} \delta_{kl} \delta_{mn}) + y_5
 (\delta_{ij} \delta_{km} \delta_{ln} \delta_{pq} + \delta_{iq} \delta_{jp}
 \delta_{km} \delta_{ln}) + y_6 (\delta_{ij} \delta_{kn} \delta_{lm} \delta_{pq} + \delta_{iq} \delta_{jp} \delta_{kn} \delta_{lm}) + \\
&y_7 (\delta_{ip}
 \delta_{jl} \delta_{km} \delta_{nq}
 + \delta_{ip} \delta_{jn} \delta_{km} \delta_{lq}) +
 y_8 (\delta_{ip} \delta_{jk} \delta_{ln} \delta_{mq} +
 \delta_{ip} \delta_{jm} \delta_{kq} \delta_{ln}) + y_9 (\delta_{ip} \delta_{jk} \delta_{lq} \delta_{mn} + \\
&\delta_{ip} \delta_{jl} \delta_{kq} \delta_{mn} +
 \delta_{ip} \delta_{jm} \delta_{kl} \delta_{nq}
 + \delta_{ip} \delta_{jn} \delta_{kl} \delta_{mq})
 + y_{10} (\delta_{ip} \delta_{jk} \delta_{lm} \delta_{nq}
 + \delta_{ip} \delta_{jl} \delta_{kn} \delta_{mq} + \delta_{ip} \delta_{jm}
 \delta_{kn} \delta_{lq} + \\
&\delta_{ip} \delta_{jn} \delta_{kq} \delta_{lm}) +
 y_{11} (\delta_{ik} \delta_{jp} \delta_{lq} \delta_{mn}
 + \delta_{ik} \delta_{jl} \delta_{mn} \delta_{pq} +
 \delta_{im} \delta_{jp} \delta_{kl} \delta_{nq} +
 \delta_{im}\delta_{jn}\delta_{kl}\delta_{pq}) + \\
&y_{12} (\delta_{il}\delta_{jp} \delta_{kq} \delta_{mn} +
 \delta_{il} \delta_{jk} \delta_{mn} \delta_{pq} +
 \delta_{in} \delta_{jp} \delta_{kl} \delta_{mq} +
 \delta_{in} \delta_{jm} \delta_{kl} \delta_{pq}) +
 y_{13} (\delta_{ik} \delta_{jp} \delta_{ln} \delta_{mq} +\\
&\delta_{ik} \delta_{jm} \delta_{ln} \delta_{pq} +
 \delta_{im} \delta_{jp} \delta_{kq} \delta_{ln} +
 \delta_{im} \delta_{jk} \delta_{ln} \delta_{pq}) +
 y_{14} (\delta_{il}\delta_{jp}\delta_{km}\delta_{nq} +
 \delta_{il}\delta_{jn} \delta_{km} \delta_{pq} +
 \delta_{in} \delta_{jp} \delta_{km} \delta_{lq} + \\
&\delta_{in}\delta_{jl}\delta_{km}\delta_{pq}) +
 y_{15} (\delta_{ik}\delta_{jp} \delta_{lm} \delta_{nq} +
 \delta_{ik} \delta_{jn} \delta_{lm} \delta_{pq} +
 \delta_{im} \delta_{jp} \delta_{kn} \delta_{lq} +
 \delta_{im} \delta_{jl} \delta_{kn} \delta_{pq}) +\\
&y_{16} (\delta_{il} \delta_{jp} \delta_{kn} \delta_{mq} +
 \delta_{il} \delta_{jm} \delta_{kn} \delta_{pq} +
 \delta_{in} \delta_{jk} \delta_{lm} \delta_{pq} +
 \delta_{in} \delta_{jp} \delta_{lm} \delta_{kq}) +
 y_{17} (\delta_{ik} \delta_{jq} \delta_{ln} \delta_{mp} +\\
&\delta_{im} \delta_{jq} \delta_{kp} \delta_{ln}) +
 y_{18} (\delta_{il} \delta_{jq} \delta_{km} \delta_{np} +
 \delta_{in} \delta_{jq} \delta_{km} \delta_{lp}) +
 y_{19} (\delta_{ik} \delta_{jq} \delta_{lm} \delta_{np} +
 \delta_{im} \delta_{jq} \delta_{kn} \delta_{lp}) +\\
&y_{20} (\delta_{il} \delta_{jq} \delta_{kn} \delta_{mp} +
 \delta_{in} \delta_{jq} \delta_{kp} \delta_{lm}) +
 y_{21} (\delta_{ik} \delta_{jq} \delta_{lp} \delta_{mn} +
 \delta_{im} \delta_{jq} \delta_{kl} \delta_{np}) +
 y_{22} (\delta_{il}\delta_{jq}\delta_{kp}\delta_{mn} +\\
&\delta_{in}\delta_{jq} \delta_{kl}\delta_{mp}) +
 y_{23} (\delta_{ik}\delta_{jl}\delta_{mq}\delta_{np} +
 \delta_{ik} \delta_{jm} \delta_{lq} \delta_{np} +
 \delta_{im} \delta_{jn} \delta_{kq} \delta_{lp} +
 \delta_{im} \delta_{jk} \delta_{lp} \delta_{nq}) +\\
&y_{24} (\delta_{il}\delta_{jk}\delta_{mq}\delta_{np} +
 \delta_{il}\delta_{jm} \delta_{kq} \delta_{np} +
 \delta_{in} \delta_{jm} \delta_{kq} \delta_{lp} +
 \delta_{in}\delta_{jk}\delta_{lp}\delta_{mq}) +
 y_{25} (\delta_{ik}\delta_{jl} \delta_{mp} \delta_{nq} + \\
&\delta_{ik} \delta_{jn} \delta_{lq} \delta_{mp} +
 \delta_{im} \delta_{jn} \delta_{kp} \delta_{lq} +
 \delta_{im} \delta_{jl} \delta_{kp}\delta_{nq}) +
 y_{26} (\delta_{ik}\delta_{jm}\delta_{lp}\delta_{nq} +
 \delta_{ik} \delta_{jn} \delta_{lp} \delta_{mq} +
 \delta_{im} \delta_{jk} \delta_{lq} \delta_{np} +\\
&\delta_{im} \delta_{jl} \delta_{kq} \delta_{np}) +
 y_{27} (\delta_{il}\delta_{jk}\delta_{mp}\delta_{nq} +
 \delta_{il} \delta_{jn} \delta_{kq} \delta_{mp} +
 \delta_{in} \delta_{jm} \delta_{kp} \delta_{lq} +
 \delta_{in} \delta_{jl} \delta_{kp} \delta_{mq}) +\\
&y_{28} (\delta_{il} \delta_{jm} \delta_{kp} \delta_{nq} +
 \delta_{il} \delta_{jn} \delta_{kp} \delta_{mq} +
 \delta_{in} \delta_{jk} \delta_{lq} \delta_{mp} +
 \delta_{in} \delta_{jl} \delta_{kq} \delta_{mp}) +
 y_{29} (\delta_{ij}\delta_{kp}\delta_{lq}\delta_{mn} +\\
&\delta_{iq} \delta_{jl} \delta_{kp} \delta_{mn} +
 \delta_{ij} \delta_{kl} \delta_{mp} \delta_{nq} +
 \delta_{iq} \delta_{jn} \delta_{kl} \delta_{mp}) +
 y_{30} (\delta_{ij}\delta_{kp}\delta_{ln}\delta_{mq} +
 \delta_{iq} \delta_{jm} \delta_{kp} \delta_{ln} +
 \delta_{ij} \delta_{kq} \delta_{ln} \delta_{mp} +\\
&\delta_{iq}\delta_{jk}\delta_{ln}\delta_{mp}) +
 y_{31} (\delta_{ij} \delta_{kp} \delta_{lm} \delta_{nq} +
 \delta_{iq} \delta_{jn} \delta_{kp} \delta_{lm} +
 \delta_{ij} \delta_{kn} \delta_{lq} \delta_{mp} +
 \delta_{iq} \delta_{jl} \delta_{kn}\delta_{mp}) +\\
&y_{32} (\delta_{ij}\delta_{kq}\delta_{lp}\delta_{mn} +
 \delta_{iq} \delta_{jk} \delta_{lp} \delta_{mn} +
 \delta_{ij} \delta_{kl} \delta_{mq} \delta_{np} +
 \delta_{iq} \delta_{jm} \delta_{kl} \delta_{np}) +
 y_{33} (\delta_{ij}\delta_{km}\delta_{lp}\delta_{nq} +\\
&\delta_{iq} \delta_{jn} \delta_{km} \delta_{lp} +
 \delta_{ij} \delta_{km} \delta_{lq} \delta_{np} +
 \delta_{iq} \delta_{jl} \delta_{km} \delta_{np}) +
 y_{34} (\delta_{ij} \delta_{kn} \delta_{lp} \delta_{mq} +
 \delta_{iq} \delta_{jm} \delta_{kn} \delta_{lp} +
 \delta_{ij} \delta_{kq} \delta_{lm} \delta_{np} +\\
&\delta_{iq} \delta_{jk} \delta_{lm} \delta_{np})
 \end{split}
\label{eq:C9}
\end{equation}

By substitution of Eqs.(\ref{eq:C6}) and (\ref{eq:C9}) into Eq.(\ref{eq:C8a}) and remembering that there are two additional terms (each of them is obtained from Eq.(\ref{eq:C9}) by a simple exchange of a pair of indices) one may get a system of $23$ equations with $34$ unknowns \footnote{The system is obtained by equating the numerical coefficients of the identical products of Kronecker deltas on both sides of the expanded version of Eq.(\ref{eq:C8a}).} which may be separated into the following four independent subsystems
\begin{alignat}{4}
&y_1 + 2 y_{29} = -b_4 & \quad &3 y_4 = -a_4 & \quad &y_5 + 2 y_{11} = -b_4
& \quad &y_2 + 2 y_{25} = -f_4 \notag \\
&y_9 + y_{29} + y_{30} = -e_4 & \quad &3 y_{16} = -d_4 & \quad
&y_{11} + y_{21} + y_{33} = -e_4 & \quad &y_7 + y_{17} + y_{25} = -j_4 \notag \\&y_9 + y_{22} + y_{31} = -e_4 & \quad &y_6 + 2 y_{12} = -c_4 & \quad
&y_{13} + y_{14} + y_{15} = -e_4 & \quad & \notag \\
&y_{10} + y_{27} + y_{30} = -i_4 & \quad &y_4 + 2 y_{32} = -c_4 & \quad
&y_5 + 2 y_{23} = -g_4 & \quad & \notag \\
&y_{10} + y_{20} + y_{28} = -i_4 & \quad &y_{12}+y_{32}+y_{34} = -d_4 & \quad
&y_{13} + y_{26} + y_{33} = -i_4 & \quad & \notag \\
&y_8 + y_{28} + y_{31} = -k_4 & \quad &y_{16} + y_{24} + y_{34} = -l_4 & \quad
&y_{14} + y_{19} + y_{23} = -i_4 & \quad & \notag \\
&y_3 + 2 y_{27} = -g_4 & \quad &y_6 +  2 y_{24} = -h_4 & \quad
&y_{15} + y_{18} + y_{26} = -k_4 & \quad & \label{eq:C19}
\end{alignat}
where the second subsystem is compatible and admits a solution if and only if
Eq.(\ref{eq:C20}) is satisfied.

\section{OBTAINING OF NUMERICAL COEFFICIENTS OF EQ.(\ref{eq:C6}).}
\label{sec:D}

Due to Eqs.(\ref{eq:C31}) and (\ref{eq:C32})
\begin{align}
a_4 + b_4 + c_4 &= 0 \label{eq:C25} \\
d_4 + i_4 + l_4 &= 0 \notag \\
b_4 + f_4 + g_4 &= -2 (b_4 + 2 e_4) \notag \\
c_4 + g_4 + h_4 &= -2 (c_4 + d_4 + e_4) \notag \\
e_4 + i_4 + k_4 &= - (e_4 + i_4 + j_4) / 2 \notag \\
\begin{split}
f_4 + 2 k_4 &= 2 (e_4 + i_4 + k_4) + (b_4 + f_4 + g_4) - (b_4 + 2 e_4) -
              (g_4 + 2 i_4) \\
            &= - (e_4 + i_4 + j_4) - 3 (b_4 + 2 e_4) - (g_4 + 2 i_4)
\end{split} \notag \\
\begin{split}
f_4 + 2 j_4 &= 2 (e_4 + i_4 + j_4) + (b_4 + f_4 + g_4) - (b_4 + 2 e_4) -
              (g_4 + 2 i_4) \\
            &= 2 (e_4 + i_4 + j_4) - 3 (b_4 + 2 e_4) - (g_4 + 2 i_4) 
\end{split} \notag \\
\begin{split}
h_4 + k_4 + l_4 &= (c_4 + g_4 + h_4) + (e_4 + i_4 + k_4) - (c_4 + d_4 + e_4) -
                  (g_4 + 2 i_4) \\
                &= -3 (c_4+d_4+e_4) - (e_4+i_4+j_4) / 2 - (g_4 + 2 i_4) 
\end{split} \notag
\end{align}
Thus, all measurable statistical moments may be expressed as combinations of $(b_4+2e_4)$, $(c_4 + d_4 + e_4)$, $(g_4 + 2 i_4)$ and $(e_4 + i_4 + j_4)$. The next step is to relate them to the components of the fourth-order statistical moments of $\mathbf{g}$ given by Eqs.(\ref{eq:C24}). This is done by means of the inversion of the following relations which are obtained by replacing Eq.(\ref{eq:C6}) in the definition of the four components $\{F_1, F_2, F_3, F_4 \}$ and, in a second step, making use of Eqs.(\ref{eq:C25}).
\begin{align}
 \begin{split}
F_1 &= (a_4 + b_4 + c_4) + 4 (b_4 + 2 e_4) + 4 (c_4 + d_4 + e_4) + (b_4 + f_4
     + g_4) + (c_4 + g_4 + h_4) + \\
    &4 (d_4 + i_4 + l_4) + 6 (e_4 + i_4 + j_4) + 6 (e_4 + i_4 + k_4) +
     2 (f_4 + 2 k_4) + 4 (g_4 + 2 i_4) + \\
    &2 (h_4+k_4+l_4) = 2 (b_4+2 e_4) + 2 (c_4+d_4+e_4) + 3 (e_4+i_4+j_4) + 
    4 (g_4+2 i_4) + \\
    &2 (f_4+2 k_4) + 2 (h_4+k_4+l_4) = -4 (b_4+2 e_4) - 4 (c_4+d_4+e_4)
 \end{split} \notag \\
 \begin{split}
F_2 &= (b_4 + 2 e_4) + (f_4 + 2 j_4) + (g_4 + 2 i_4) + 2 (e_4 + i_4 + k_4)
     = (b_4 + 2 e_4) + (g_4 + 2 i_4) - \\
    &(e_4 + i_4 + j_4) + (f_4 + 2 j_4)
      = - 2 (b_4 + 2 e_4) + (e_4 + i_4 + j_4)
 \end{split} \label{eq:C75} \\
F_3 &= 3 (f_4 + 2 j_4) = 6 (e_4 + i_4 + j_4) - 9 (b_4 + 2 e_4) -
3 (g_4 + 2 i_4) \notag \\
F_4 &= (b_4 + f_4 + g_4) = - 2 (b_4 + 2 e_4) \notag
\end{align}
This inversion gives rise to Eqs.(\ref{eq:C27}).

\section{DERIVATION OF EQS.(\ref{eq:C24}).}
\label{sec:E}

In order to derive Eqs.(\ref{eq:C24}), one must compute the values of $\{I_1, I_2, I_3, I_4 \}$ and $\{F_1, F_2, F_3, F_4 \}$ and, then, compare them.

In the first derivation of these equations~\cite{Siggia:1981}, the calculation of the values of $\{I_1, I_2, I_3, I_4 \}$ was very easy from the generic expressions of the fourth-order correlations of $\mathbf{g}$ as functions of strain and vorticity components. However, the calculation of $\{F_1, F_2, F_3, F_4 \}$ was not so easy, since it involved the expression of each generic component of $\mathbf{g}$ as the addition of a symmetric and an antisymmetric part and the application of the distributive property to the resulting fourth-order products of additions of symmetric and antisymmetric parts. Finally, one had to compute each resulting term, from the known generic expressions of the fourth order correlations, and had to add them. In fact, these calculations were indicated but not developed~\cite{Siggia:1981}.

In this paper, the approach is the opposite. $\{F_1, F_2, F_3, F_4 \}$ are computed straightforwardly from Eq.(\ref{eq:C6}), whereas $\{I_1, I_2, I_3, I_4 \}$ need more work.

The values of $\{F_1, F_2, F_3, F_4 \}$ have already been computed in Appendix~\ref{sec:D}. Eqs.(\ref{eq:C75}) relate them to the measurable combinations of coefficients of Eq.(\ref{eq:C6}). On the other hand, the values of $\{I_1, I_2, I_3, I_4 \}$ are obtained from their definitions.

\begin{align}
I_1 &= \langle (tr(\mathbf{s^2}))^2 \rangle = \langle s_{ij} s_{ji} s_{mn} s_{nm} \rangle = \frac{\delta_{il} \delta_{jk} \delta_{mq} \delta_{np}}{16} \langle (g_{ij} + g_{ji}) (g_{kl} + g_{lk}) (g_{mn} + g_{nm}) (g_{pq} + g_{qp}) \rangle
\notag \\
I_2 &= \langle \omega^2 tr(\mathbf{s^2}) \rangle = \langle \omega_u \omega_u s_{mn} s_{nm} \rangle = \frac{\epsilon_{uij} \epsilon_{ukl} \delta_{mq} \delta_{np}}{4} \langle g_{ij} g_{kl} (g_{mn} + g_{nm}) (g_{pq} + g_{qp}) \rangle
\label{eq:C76} \\
I_3 &= \langle \omega_i s_{ij} s_{jl} \omega_l \rangle = \frac{\epsilon_{imn} \epsilon_{lpq} \delta_{jk}}{4} \langle (g_{ij} + g_{ji}) (g_{kl} + g_{lk}) g_{mn} g_{pq} \rangle 
\notag \\
I_4 &= \langle (\omega^2)^2 \rangle = \langle \omega_u \omega_u \omega_v \omega_v \rangle = \epsilon_{uij} \epsilon_{ukl} \epsilon_{vmn} \epsilon_{vpq} \langle g_{ij} g_{kl} g_{mn} g_{pq} \rangle
\notag
\end{align} 

In Eqs.(\ref{eq:C76}), $\epsilon_{imn}$ is the Levi-Civitta symbol whose value is either $+1$ or $-1$ or $0$, depending on whether $imn$ is an even permutation of $\{1,2,3\}$, an odd permutation or there are some repeated indices. Because of the definition of the Levi-Civitta symbol, one has
\begin{equation*}
\epsilon_{imn} \epsilon_{lpq} = \delta_{il} \delta_{mp} \delta_{nq} - \delta_{il} \delta_{mq} \delta_{np} + \delta_{ip} \delta_{mq} \delta_{nl} - \delta_{ip} \delta_{ml} \delta_{nq} + \delta_{iq} \delta_{ml} \delta_{np} - \delta_{iq} \delta_{mp} \delta_{nl}
\end{equation*}
This expression should be used wherever there is a product of two Levi-Civitta symbols. Next, an interesting particular case of the previous expression is worked out
\begin{equation*}
\epsilon_{uij} \epsilon_{ukl} = \delta_{ik} \delta_{jl} - \delta_{il} \delta_{jk}
\end{equation*}

The application of Eq.(\ref{eq:C6}) to the definition of $I_1$ in Eqs.(\ref{eq:C76}) gives
\begin{equation}
I_1 = 9 a_4 + 48 (b_4 + c_4 + d_4 + 3 e_4 + f_4 + 2 g_4 + h_4) + 57 (4 i_4 + j_4 + 2 k_4 + l_4)
\label{eq:C77}
\end{equation}
Eqs.(\ref{eq:C31}) and (\ref{eq:C32}) imply that $$a_4 + b_4 + c_4 = d_4 + i_4 + l_4 = 3 e_4 + 3 i_4 + j_4 + 2 k_4 = 0$$ So that Eq.(\ref{eq:C77}) may be rewritten as
\begin{equation*}
I_1 = 48 (b_4 + f_4 + g_4) + 48 (c_4 + g_4 + h_4) + 9 (4 i_4 + j_4 + 2 k_4 + l_4 - b_4 - c_4)
\end{equation*}
Adding and subtracting $d_4 + 3 e_4$ in the last term to the right of the previous expression, one gets
\begin{equation*}
I_1 = 48 (b_4 + f_4 + g_4) + 48 (c_4 + g_4 + h_4) - 9 (b_4 + 2 e_4) - 9 (c_4 + d_4 + e_4)
\end{equation*}
Eqs.(\ref{eq:C32}) and (\ref{eq:C27}) may be applied now to obtain
\begin{equation}
I_1 = -105 [(b_4 + 2 e_4) + (c_4 + d_4 + e_4)] = \frac{105}{4} F_1
\label{eq:C78}
\end{equation}

Developing the definition of $I_2$ in Eqs.(\ref{eq:C76}), one may get
\begin{equation}
I_2 = \frac{\delta_{ik} \delta_{jl} \delta_{mq} \delta_{np} + \delta_{ik} \delta_{jl} \delta_{mp} \delta_{nq} - \delta_{il} \delta_{jk} \delta_{mq} \delta_{np} - \delta_{il} \delta_{jk} \delta_{mp} \delta_{nq}}{2} \langle g_{ij} g_{kl} g_{mn} g_{pq} \rangle
\label{eq:C79}
\end{equation}
Although Eq.(\ref{eq:C79}) should have contained the addition of $8$ terms with a divisor $4$, the remaining terms were equal, under commutativity of the components of $\mathbf{g}$, to those which do appear explicitly in Eq.(\ref{eq:C79}). Therefore, their contribution was a factor $2$ which was simplified with the divisor $4$. It should also be noticed that the first and the last product of Kronecker deltas in the right hand side of Eq.(\ref{eq:C79}) share the same numerical coefficient, $g_4$, in Eq.(\ref{eq:C6}). It means that they are equal under commutativity and their contributions cancel each other because they have opposite sign. Finally, one must apply Eq.(\ref{eq:C6}) to calculate the simplified expression where there are only two remaining terms
\begin{equation}
\begin{split}
I_2 = &\frac{\delta_{ik} \delta_{jl} \delta_{mp} \delta_{nq} - \delta_{il} \delta_{jk} \delta_{mq} \delta_{np}}{2} \langle g_{ij} g_{kl} g_{mn} g_{pq} \rangle =\\
&6 [3 (b_4 - c_4) - 4 (d_4 - e_4) + 6 (f_4 - h_4) + 9 j_4 - 2 k_4 - 7 l_4]
\end{split} \label{eq:C80}
\end{equation}
By using Eqs.(\ref{eq:C28}) in Eq.(\ref{eq:C80}), one gets
\begin{equation}
I_2 = -\frac{45}{2} F_1 + 60 F_2 + 30 F_4
\label{eq:C81}
\end{equation}

The development of the definition of $I_3$ in Eqs.(\ref{eq:C76}) would produce an addition of $24$ contributions when carried out. However, most of them cancel each other because they are equal under commutativity of components of $\mathbf{g}$ and have opposite sign. In the end, there are only four contributions which do not cancel and must be computed
\begin{equation}
I_3 = \frac{(\delta_{ik} \delta_{jl} \delta_{mp} \delta_{nq} - \delta_{il} \delta_{jk} \delta_{mq} \delta_{np}) + (\delta_{il} \delta_{jp} \delta_{kn} \delta_{mq} - \delta_{ik} \delta_{jq} \delta_{ln} \delta_{mp})}{2} \langle g_{ij} g_{kl} g_{mn} g_{pq} \rangle
\label{eq:C82}
\end{equation}
The application of Eq.(\ref{eq:C6}) to this Eq.(\ref{eq:C82}) gives,
\begin{equation}
\begin{split}
I_3 = &6 [3 (b_4 - c_4) - 4 (d_4 - e_4) + 6 (f_4 - h_4) + 9 j_4 - 2 k_4 - 7 l_4] - \\
&6 [2 (b_4 - c_4) - 6 (d_4 - e_4) + 4 (f_4 - h_4) + 11 j_4 - 3 k_4 - 8 l_4] =\\
&6 [(b_4 - c_4) + 2 (d_4 - e_4) + 2 (f_4 - h_4) - 2 j_4 + k_4 + l_4]
\end{split} \label{eq:C83}
\end{equation}
By using Eqs.(\ref{eq:C28}) in Eq.(\ref{eq:C83}), one gets
\begin{equation}
I_3 = -\frac{15}{2} F_1 - 15 F_2 + 45 F_4
\label{eq:C84}
\end{equation}

The development of the definition of $I_4$ in Eqs.(\ref{eq:C76}) gives
\begin{equation}
I_4 = (\delta_{ik} \delta_{jl} \delta_{mp} \delta_{nq} - \delta_{ik} \delta_{jl} \delta_{mq} \delta_{np} - \delta_{il} \delta_{jk} \delta_{mp} \delta_{nq} + \delta_{il} \delta_{jk} \delta_{mq} \delta_{np}) \langle g_{ij} g_{kl} g_{mn} g_{pq} \rangle
\label{eq:C85}
\end{equation}
In Eq.(\ref{eq:C85}) the second and the third term in the right hand side correspond to the same contribution, under commutativity of the components of $\mathbf{g}$, with the same sign; so, they do not cancel each other, but it will be sufficient to compute one of them with a factor $2$. The application of Eq.(\ref{eq:C6}) to Eq.(\ref{eq:C85}) gives,
\begin{equation}
\begin{split}
I_4 = &60 f_4 - 120 g_4 + 60 h_4 - 240 i_4 + 60 j_4 + 120 k_4 + 60 l_4 = \\
&30 (f_4 + 2 j_4) + 30 (f_4 + 2 k_4) + 60 (h_4 + k_4 + l_4) - 120 (g_4 + 2 i_4)
\end{split} \label{eq:C86}
\end{equation}
Using Eqs.(\ref{eq:C28}) in Eq.(\ref{eq:C86}), one gets
\begin{equation}
I_4 = 45 F_1 - 480 F_2 + 80 F_3 + 120 F_4
\label{eq:C87}
\end{equation}

Eqs.(\ref{eq:C78}), (\ref{eq:C81}), (\ref{eq:C84}) and (\ref{eq:C87}) make up a system of equations which is, exactly, the inverse of Eqs.(\ref{eq:C24}). The precedent statement is better appreciated if the full system is written in matrix form
\begin{equation}
\begin{pmatrix}
I_1 \\ I_2 \\ I_3 \\ I_4
\end{pmatrix}
=
\begin{pmatrix}
105 / 4 & 0 & 0 & 0 \\
-45 / 2 & 60 & 0 & 30 \\
-15 / 2 & -15 & 0 & 45 \\
45 & -480 & 80 & 120
\end{pmatrix}
\begin{pmatrix}
F_1 \\ F_2 \\ F_3 \\ F_4
\end{pmatrix}
\label{eq:C88}
\end{equation}
and compared to Eqs.(\ref{eq:C24}), rewritten in matrix form,.
\begin{equation*}
\begin{pmatrix}
F_1 \\ F_2 \\ F_3 \\ F_4
\end{pmatrix}
=
\begin{pmatrix}
4 / 105 & 0      & 0        & 0 \\
1 / 105 & 1 / 70 & -1 / 105 & 30 \\
3 / 140 & 11 / 140 & -3 / 35 & 1 / 80 \\
1 / 105 & 1 / 210 & 2 / 105 & 0
\end{pmatrix}
\begin{pmatrix}
F_1 \\ F_2 \\ F_3 \\ F_4
\end{pmatrix}
\end{equation*}

\newpage
\bibliographystyle{unsrt}
\bibliography{biblio}

\end{document}